# Cooperative Control in Production and Logistics[*]

**László Monostori[1,2,**], Paul Valckenaers[3], Alexandre Dolgui[4], Hervé Panetto[5], Mietek Brdys[6,7], Balázs Csanád Csáji[1]**

[1] *Institute for Computer Science and Control, Hungarian Academy of Sciences, Hungary*
[2] *Faculty of Mechanical Engineering, Budapest University of Technology and Economics, Hungary*
[3] *Department of Health Care and Technology, KHLeuven, Belgium*
[4] *LIMOS UMR 6158 CNRS, Ecole Nationale Supérieure des Mines de Saint-Etienne, France*
[5] *Centre de Recherche en Automatique de Nancy (UMR 7039), Université de Lorraine, CNRS, France*
[6] *School of Electronic, Electrical and Computer Engineering, University of Birmingham, United Kingdom*
[7] *Department of Control Systems Engineering, Gdansk University of Technology, Gdansk, Poland*

**Abstract:** Classical applications of control engineering and information and communication technology (ICT) in production and logistics are often done in a rigid, centralized and hierarchical way. These inflexible approaches are typically not able to cope with the complexities of the manufacturing environment, such as the instabilities, uncertainties and abrupt changes caused by internal and external disturbances, or a large number and variety of interacting, interdependent elements. A paradigm shift, e.g., novel organizing principles and methods, is needed for supporting the interoperability of dynamic alliances of agile and networked systems. Several solution proposals argue that the future of manufacturing and logistics lies in network-like, dynamic, open and reconfigurable systems of cooperative autonomous entities.

The paper overviews various distributed approaches and technologies of control engineering and ICT that can support the realization of cooperative structures from the resource level to the level of networked enterprises. Standard results as well as recent advances from control theory, through cooperative game theory, distributed machine learning to holonic systems, cooperative enterprise modelling, system integration, and autonomous logistics processes are surveyed. A special emphasis is put on the theoretical developments and industrial applications of Robustly Feasible Model Predictive Control (RFMPC). Two case studies are also discussed: i) a holonic, PROSA-based approach to generate short-term forecasts for an additive manufacturing system by means of a delegate multi-agent system (D-MAS); and ii) an application of distributed RFMPC to a drinking water distribution system.

## 1. INTRODUCTION

The development in control engineering and information and communication technology (ICT) always acted as important enablers for newer and newer solutions – moreover generations – in production and logistics.

As to discrete manufacturing, developments in ICT led to the realization of product life-cycle management (PLM), computer numerical control (CNC), enterprise resource planning (ERP) and computer integrated manufacturing (CIM) systems. Integration often resulted in rigid, centralized or hierarchical control architectures which could not cope with an unstable and uncertain manufacturing environment: internal as well as external disturbances in manufacturing and related logistics and frequently changing market demands.

Growing complexity is another feature showing up in production and logistics processes, furthermore, in enterprise structures, as well (Wiendahl and Scholtissek, 1994; Schuh, *et al.*, 2008; ElMaraghy, *et al.*, 2012). Decision should be based on the pertinent information; time should be seriously considered as a limiting resource for decision-making, and the production and logistics systems should have changeable, easy-to-reconfigure organizational structures.

New organizing principles and methods are needed for supporting the interoperability of dynamic virtual alliances of agile and networked systems which – when acting together – can make use of opportunities without suffering from diseconomies of scale (Monostori, *et al.*, 2006).

Various solution proposals unanimously imply that the future of manufacturing and logistics lies in the loose and temporal federations of cooperative autonomous entities (Vámos, 1983). The interaction of individuals may lead to emergence of complex system-level behaviors (Ueda *et al.*, 2001). Evolutionary system design relies on this emergence when modelling and analyzing complex manufacturing and logistics in a wider context of eco-technical systems.

---

Under the pressure of the challenges highlighted above, the transformations of manufacturing and logistics systems are already underway (Jovane *et al.*, 2003). The need for novel organizational principles, structures and method has called for various approaches (Tharumarajah, *et al.*, 1996) in the past decades, such as holonic (Van Brussel, *et al.*, 1998; Valckenaers and Van Brussel, 2005), fractal (Warnecke, 1993), random (Iwata *et al.,* 1994), biological (Ueda, *et al.*, 1997), multi-agent manufacturing systems (Bussmann, *et al.*, 2004; Monostori, *et al.*, 2006), bucket brigades (Bartholdi and Eisenstein, 1996; Bratcu & Dolgui, 2005; Dolgui and Proth, 2010), and autonomous logistics systems (Scholz-Reiter and Freitag, 2007).

In a milestone paper (Nof, *et al.*, 2006) – based on the scopes, activities and results of all the Technical Committees (TCs) of the Coordinating Committee on Manufacturing and Logistics Systems (CC5) of the International Federation of Automatic Control (IFAC) – four emerging trends for solution approaches were identified (Figure 1).

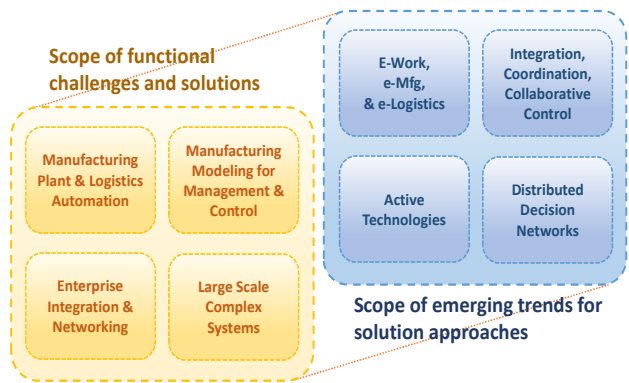

Fig. 1. Scope of functional challenges/solutions and emerging trends for solution approaches (Nof, *et al.*, 2006).

The aforementioned milestone paper, concentrating on e-work, e-manufacturing and e-logistics enabled by the Internet, underlined the importance of understanding how to model, design and control effective e-work, in order to secure the productivity and competitiveness of manufacturing and logistics systems.

In addition to cooperativeness, another indispensable characteristic of production and logistics systems of the future, namely responsiveness, was underlined in (Váncza *et al.*, 2011) where the concept of *cooperative and responsive manufacturing enterprises (CoRMEs)* was introduced and the heavy challenges in their realization were emphasized together with some possible resolutions of them.

Applications of cooperative control approaches – as in many fields – raise real, difficult to answer challenges in manufacturing and logistics. These challenges – see, for example, Figure 2 – are heavy because they are directly stemming from generic conflicts between competition and cooperation, local autonomy and global behavior, design and emergence, planning and reactivity, as well as uncertainty and abundance of information (Váncza *et al.*, 2011).

- Global performance
- Cooperation
- Information sharing
- Communication
- Adaptivity
- Responsiveness
- Optimization
- Expected behaviour
- Mass production's efficiency
- Local autonomy
- Competition
- Private data, security
- Truthfulness, reputation
- Network stability
- Economic production
- Robustness
- Emergent properties
- Customization

Fig. 2. Compelling challenges of cooperative production and logistic systems.

Advantages (e.g., why should cooperative control approaches be used in production and logistics) include

- *Openness* (e.g., easier to build and change)
- *Reliability* (e.g., fault tolerance)
- *Performance* (e.g., distributed execution of tasks)
- *Scalability* (e.g., the potential of addressing large-scale problems, incremental design)
- *Flexibility* (e.g., redesign, heterogeneity)
- *Cost* (e.g., potential cost reductions)
- *Distribution* (e.g., natural for spatially separated units)

While some disadvantages of cooperative control systems, which need to be addressed, are as follows

- *Communication Overhead* (e.g., time / cost of sharing information)
- *Decentralized Information* (e.g., local vs global data)
- *Security / Confidentiality* (are harder to guarantee)
- *Decision "Myopia"* (e.g., local optima)
- *Chaotic Behavior* (e.g., butterfly effects, bottlenecks)
- *Complex to Analyze* (compared to centralized systems)

The main aim of the paper is to highlight how distributed control approaches can contribute at least to partially reduce the disadvantages while using completely the advantages, i.e. to find a safe – sometimes even narrow – path in between two extremes (only advantages or disadvantages).

Another goal of the paper is to survey distributed methods of control theory and ICT which can support the realization of cooperative structures from the resource level to the level of networked enterprises (top right circle of Figure1).

The structure of the paper is as follows. Section 2 overviews a few relevant concepts from the theory of distributed-, multi-agent-based and cooperative control systems. Section 3 aims at discussing recent approaches to Robustly Feasible Model Predictive Control, which is one of the highlights of the paper. Later, Section 4 presents some existing paradigms and specialized cooperative technologies designed and applied in production and logistics. Two case studies are presented in Section 5, a holonic approach to generate forecasts for additive manufacturing, and an application of RFMPC to water management. Finally, Section 6 concludes the paper.

## 2. DISTRIBUTED, AGENT-BASED AND COOPERATIVE CONTROL APPROACHES

Classical control theory (Glad and Ljung, 2000; Åström and Murray, 2008) usually aims at designing a controller, namely, a decision making unit with limited processing capacities, which interacts with a (typically uncertain, dynamic) system.

There are several ways to model the object to be controlled, from simple linear transfer functions, rational maps and state space models, to even nonparametric, nonlinear models, such as neural networks, kernel machines, wavelets and fuzzy systems (Ljung, 1999). Basic concepts, such as long-term costs, sensitivity and stability are often applied performance indicators to measure the quality of the controller.

Results of classical control theory are widely applied in various fields of production (Chryssolouris, 2006) and logistics (Ivanov et al., 2012; Song, 2013).

### 2.1 Distributed Control

On the other hand, classical results typically focus on a single controller, while in practice there are usually several decision making units which interact with each other based on limited inter-component communications (Shamma, 2007). These interactions are crucial and should also be taken into account when designing complex production and logistics systems.

In a *distributed control system* there are more (not necessarily autonomous) decision making units which can operate in parallel and typically control various sub-systems of a complex system. The controllers are interconnected, usually monitor and communicate with each other via a network and often regulated by a central controller (Meyn, 2007).

One of the basic principles of distributed control is to divide a complex control task into several smaller ones which can be addressed by local control units that are simpler to design and operate. This idea is often called *divide-and-conquer*, and it typically also speeds up the computation as calculating the sub-solutions can be often done in a distributed way (Wu et al., 2005). It is a key issue, as well, that such systems are *modular* and hence more *robust* (Perkins et al., 1994).

### 2.2 Multi-Agent Paradigm

A *Multi-Agent System* (MAS) can be both viewed as a special type of localized distributed control system of autonomous control units as well as a novel systemic paradigm to organize humans and machines as a whole system.

An *agent* is basically a *self-directed* entity with its own value system and a means to communicate with other such objects (Baker, 1998). It archetypally makes local decisions based mainly on locally available, usually partial information. The limited information and processing power of agents are often emphasized with the term *bounded rationality*. Agents may represent any entity with self-orientation, such as cells, species, individuals, vehicles, machines, firms or nations.

The interaction between the agents can be *active*, e.g., direct message sending, or *passive*, for example, they have access to and influence the same object of the environment.

A MAS, especially with a heterarchical architecture, can show up several advantages (Baker, 1998), such as self-configuration, scalability, fault tolerance, massive parallelism, reduced complexity, increased flexibility, reduced cost and emergent behavior (Ueda et al., 2001).

A MAS approach could be useful for enterprises which often need to change their configurations (factories, inventories, fleets, etc.) by adding or removing resources; enterprises for which it is hard to predict the possible scenarios according to which they will need to work in the future (Baker, 1998).

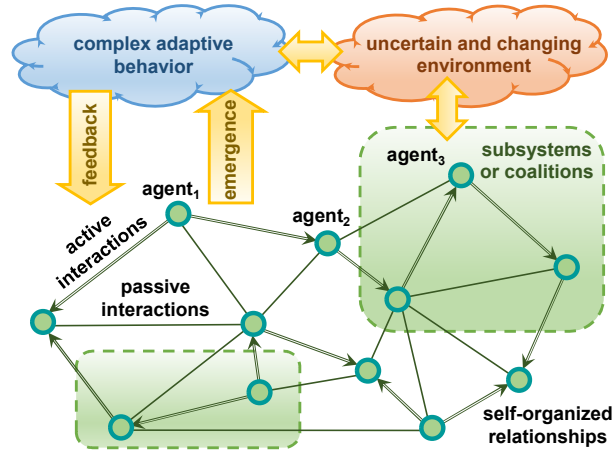

Fig. 3. The emergence of a complex adaptive behavior via interactions of the agents with their environment.

One of the key properties of an agent is its capacity to *learn* and *adapt* to its environments. A *Complex Adaptive System* (CAS) can be considered as a MAS with highly adaptive agents (Holland, 1992, 1995). Environmental conditions are changing due to the agents' interactions as they compete and cooperate for the same resources or for achieving a specific goal. This changes the behavior of the agents themselves, as well. The most remarkable phenomenon exhibited by a CAS is the *emergence* of highly structured collective behavior over time from the interactions of simple subsystems. The emergence of a global behavior is illustrated by Figure 3.

Multi-agent based or *holonic* manufacturing systems with adaptive agents received a great deal of recent attention (Márkus et al., 1996; Baker, 1998; Monostori et al., 2006; Schuh et al., 2008; Váncza et al., 2011). They became an important tool for managing various forms of complexity and optimizing diverse types of production and logistic systems.

Many complex adaptive system models were inspired by *biological systems* (Kennedy and Eberhart, 1995), such as bird flocks, wolf packs, fish schools, termite hills or ant-colonies. These approaches can show up strongly robust and parallel behavior. On the other hand, they often have the disadvantage that they are hard to theoretically analyze, for example, predicting the behavior of the system in case of various scenarios is challenging.

Agent-based *simulation* is a practical way of addressing the issue of hard theoretical analysis. Simulation became one of the standard tools to investigate the long-term behavior of MASs and to test their responses to various scenarios.

There are several modelling frameworks and semi-formal languages available to design MAS based systems, including ASRM: *Agent Systems Reference Model* (Regli et al., 2009); DAML: *DARPA Agent Markup Language* (Berners-Lee et al., 2001); EMAA: *An Extendable Mobile Agent Architecture* (Lentini et al., 1998); AML: *Agent Modelling Language* (Trencansky and Cervenka, 2005); CLAIM: *Computational Language for Autonomous, Intelligent and Mobile Agents* (Fallah-Seghrouchni and Suna, 2004) and AUML: *Agent Unified Modelling Language* (Haugen and Runde, 2008; www.auml.org) which is an initiative of FIPA: *Foundation for Intelligent Physical Agents* (www.fipa.org) which is itself an IEEE *Computer Society* standards organization.

One of such frameworks is the so-called PROSA: *Product Resource Order Staff Agents* (Van Brussel et al., 1998) reference architecture which was designed especially for MASs in production and logistics (see also Section 4.2).

*2.3 Cooperative Control*

While in a multi-agent system the agents may compete for the limited resources, e.g., the loss of one agent can be a gain for another one, in a *Cooperative Control System* (CCS) the entities should collaborate to achieve a common goal, which typically none of them could achieve by itself.

A *cooperative system* (Grundel et al., 2007) usually contains (a) *more than one* decision making units; (b) the decisions of the units influence a *common decision space*; (c) the decision makers share at least one *common objective*; and (d) the entities *share information* either actively or passively.

Typical additional features of a CCS (Shamma, 2007) are (e) the *distribution of information*, as usually none of the agents have access to all of the information the other agents have gathered even if they share; and (f) *complexity*, namely, even if all the information were available, the inherent complexity of the problem often prohibits centralized solutions, hence, a divide-and-conquer type of approach is preferred.

An archetypical example of a cooperative control system is a fleet of *unmanned autonomous vehicles* with common goals, such as rendezvous, achieving a specific formation, coverage or reaching a target location (Shamma, 2007). For example, automated forklifts may self-organize to provide an efficient service for machines in a shop floor.

*2.3.1 Cooperation in Control Theory*

Many concepts and results of classical control theory can be extended to the case of several cooperating controllers. One of such fundamental notions, to which several other control theoretical concepts can also be deduced, is *stability*. Here, we start our discussion with stability of distributed systems.

There are several possible viewpoints on stability, such as (Lyapunov) stability, asymptotic stability, global asymptotic stability, and input-to-state stability (Nof, 2009). It is well-known that interconnecting *stable* systems can result in an *unstable* global system behavior. Hence, the global stability of a system is a stronger concept than the local stability of subsystems. Standard approaches to handle this problem include *small-gain theorems*, which are generalizations of the *Nyquist criterion*. They typically deal with two systems interconnected in a feedback-loop. This provides sufficient conditions for their joint stability, e.g., the interconnected system is input-to-state-stable (ISS) if the composition of specific class functions of the interconnected subsystems is a contraction (Nof, 2009). Small-gain theorems can be extended to networks of inter-connected systems and to weaker stability concepts, such as *integral* input-to-state-stability (Ito et al., 2013).

Control of complex networks became an active research area which extended several classical concepts, such as queuing, workload control, safety-stocks control via communication channels, and networked systems (Meyn, 2007).

Another classical approach with distributed generalization is *Model Predictive Control* (MPC) or *receding horizon control* (Rawlings and Mayne, 2009) which is a widespread technique with several industrial applications (Qin and Badgwell, 2003), especially in chemical plants, utilities, mining, metallurgy, food processing and power systems.

MPC relies on a (often, but not necessarily linear) dynamic model of the environment, which can be estimated from experimental data, e.g., by system identification methods (Ljung, 1999), and computes an optimal strategy (w.r.t. the identified model and a given usually linear or quadratic criterion) for a *finite time horizon*. It applies the computed control for the current time-window, and re-computes the controller based on the feedbacks for a shifted horizon.

Distributed variants of MPC often decompose the system into several sub-problems and every instance is associated with a dedicated agent. The aim of such decomposition is twofold: (1) to ensure reducing the problem size and (2) these sub-problems should have only few common decision variables. Each agent tries to solve its own sub-problem, while the agents iteratively cooperate to exchange information about their shared decision variables (Camponogara et al., 2002).

*2.3.2 Cooperative Games and Consensus Seeking*

Even classical *game theory* (von Neumann and Morgenstern, 1953) has concepts which are widely used in distributed systems design, such as zero-sum games and Nash-equilibria. The theory of sequential and cooperative games (Branzei et al., 2008) are even more relevant to CCSs, however, many important concepts, such as mechanism design, bargaining, coalition theory, and correlated equilibrium are not widely known by CCS experts, yet (Shamma, 2007). Still, there are several successful applications of game theoretical concepts for handling cooperative control problems (Shamma, 2013) and their applications in logistics (Dolgui and Proth, 2010).

Here some basic concepts of game theory, which are often used in cooperative control systems, are recalled. Only games with *transferable utilities* (TU games) are considered. In a TU game the players can form *coalitions* and it is assumed that the coalitions can divide their worth in any possible way among its members (Peleg and Sudhölter, 2004), namely, every feasible payoff is possible.

Cooperation can be modelled in various ways. Games are with *crisp coalitions* if each agent is either fully part of a coa-lition or it is not. On the other hand, in a game with *fuzzy coalitions*, several participation levels are allowed. An example for a situation where fuzzy coalitions are useful is a joint project in which the participants have some private resources (such as commodities, time, and money) and have to decide about the amount invested (Branzei *et al.*, 2008).

The *Shapley value* (Branzei *et al.*, 2008) is one of the basic one-point solution concepts of cooperative game theory, often used to evaluate the surplus generated by the coalitions. One interpretation of the Shapely value for a player is that it shows his marginal contribution to the coalitions. The application fields of Shapley value are broad, they include general resource and cost allocation (Hougaard, 2009), power transmission planning (Yen *et al.*, 1998), and sequencing and queuing (Aydinliyim and Vairaktarakis, 2011).

The concept of *consensus seeking* (Blondel *et al.*, 2005) became one of the standard ways of addressing some cooperative control problems and also often used in MASs to achieve self-organization. A *consensus protocol* is basically an interaction rule that specifies the information exchange between the agents. During consensus seeking the agents communicate using a specified protocol via a communication network. This results in changing their behavior, which is often described by an *opinion dynamics* (Olfati-Saber, 2007). The disagreement of the participants at a given time is typically modelled with a potential function. Consensus is reached if the opinion dynamics of the agents reach equilibrium. There are several theorems available about various consensus protocols, such as the *Average-Consensus Theorem* by Olfati-Saber and Murray (2004) for *linear* ones, which even guarantees exponentially fast convergence to a consensus under some special conditions about the communication network (e.g., its directed graph is balanced).

Large number of mobile agents, sometimes called as swarms, are typically governed by consensus seeking protocols. These agent groups can be used to gather and distribute resources, e.g., goods and information. Some of their applications are surveillance, search and rescue and disaster relief (Olfati-Saber, 2007). Flocking agents are typically governed by consensus algorithms. For example, they should align their velocities and directions, avoid colliding to each other and to obstacles, keep cohesion by staying within a specified radius, and reach a target or explore an area.

Some of the recent advances of consensus seeking include nonlinear consensus protocols, consensus with quantized states, consensus on random graphs, ultrafast consensus and consensus using potential games (Olfati-Saber, 2007).

Typical applications of consensus seeking protocols include formation flight of unmanned air vehicles, e.g., synchronizing heading angles, velocities, or positions (Shamma, 2007), timing, rendezvous, flocking in swarm control problems (Blondel *et al.*, 2005), as well as to manage clusters of satellites, communication networks and even automated highway systems (Olfati-Saber and Murray, 2003).

Challenges of the consensus seeking paradigm (Shamma, 2007) include: (a) *strategic decision-making*, determining, coordinating and executing a higher-level cooperation plan;
(b) construction of *datasets of benchmark scenarios*, which would help comparing various CCS approach.

### 2.3.3 Cooperative Learning

The ability to learn how to perform task effectively and to adapt to environmental changes are key issues for agents, in order to achieve efficient global system behavior. The field of machine learning classically aims at designing algorithms and data structures which allow agents to learn and adapt either using direct feedbacks or the experience of their own results.

*Machine Learning* (ML) is divided into 3 main paradigms, namely: (a) *supervised learning* (such as neural networks, kernel machines, and Bayes classifiers); (b) *self-organized* or *unsupervised learning* (such as clustering, feature extraction, and Kohonen maps); and (c) *reinforcement learning* (such as temporal difference learning, Q-learning and SARSA).

The area of distributed and parallel approaches to ML has been an active research domain since decades. One of the standard problems is to scale up classical learning algorithms to huge problems in presence of distributed information (Bekkerman *et al.*, 2012). It is beyond the scope of the paper to give an exhaustive overview about such cooperative ML approaches, only some of them, which were already applied to production and logistics problems, are highlighted.

In the standard paradigm of *Reinforcement Learning* (RL) an agent interacts with a stochastic environment. In each step, an agent makes an action and then receives both the new state of the environment and an immediate reward. The consequences of actions may only realize much later. RL aims at finding an optimal control policy which maximizes the agent's rewards on the long run (Sutton and Barto, 1998).

*Swarm optimization* methods were inspired by various biological systems. They are very robust, can naturally adapt to disturbances and environmental changes. A classic example is the ant-colony optimization algorithm (Moyson and Manderick, 1988) which is a distributed and randomized algorithm to solve shortest path problems in graphs. The ants continuously explore the current situation and the obsolete data simply *evaporates* if it is not refreshed regularly, like the pheromone in the guiding analogy of food-foraging ants.

The PROSA architecture can also be extended by ant-colony type optimization methods (Hadeli *et al.*, 2004). The main assumption is that the agents are much faster than the ironware that they control, and that makes the system capable to forecast, i.e., they can emulate the behavior of the system several times before the actual decision is taken.

A closely related concept is *Particle Swarm Optimization* (PSO) in which several candidate solutions, "particles", are maintained which explore the search space in a cooperative way. PSO was applied, e.g., for optimizing production rate and workload smoothness by Akyol and Bayhan (2011).

### 3. COOPERATIVE ROBUSTLY FEASIBLE MODEL PREDICTIVE CONTROL

A key property of Model Predictive Control (MPC), is its capacity of satisfying the constraints imposed on the control inputs, states and controlled outputs under uncertain disturbance inputs and structural and parameter uncertainties in the plant dynamics model. This is known as robust feasibility and the MPC related technology is known as Robustly Feasible Model Predictive Control (RFMPC). As MPC based controllers are already widely applied in industry, they also has the potential of controlling cooperative structures. Thus, this section presents the theory RFMPC and its applications in cooperative control design.

*3.1 Robustly Feasible Model Predictive Controller*

There are several approaches to design a *Robustly Feasible Model Predictive Controller* (RFMPC). A robust control invariant set can be determined for the MPC control law based on its nominal model and the uncertainty bounds so that if the initial state belongs to this set the recursive robust feasibility is guaranteed (Kerrigan and Maciejowski, 2001; Grieder et al., 2003). Constructive algorithms were produced to determine such sets for linear dynamic systems under the additive and polytopic set bounded uncertainty models. Safe feasibility tubes in the state space were designed and utilized to synthesize RFMPC (Langson et al., 2004; Mayne et al., 2005). A reference governor approach was proposed and investigated (Bemporad and Mosca, 1998; Angeli, et al., 2001). It was also studied for the tracking problem (Bemporad et al., 1998), where a reference trajectory over prediction horizon is designed with extra constraints being imposed during the reference trajectory generation. The calculated control inputs under the on-line updated reference trajectory can maneuver the system to the desired states without violating the state constraints under all possible uncertainty scenarios. The additional constraints on the reference are calculated based on the uncertainty bounds.

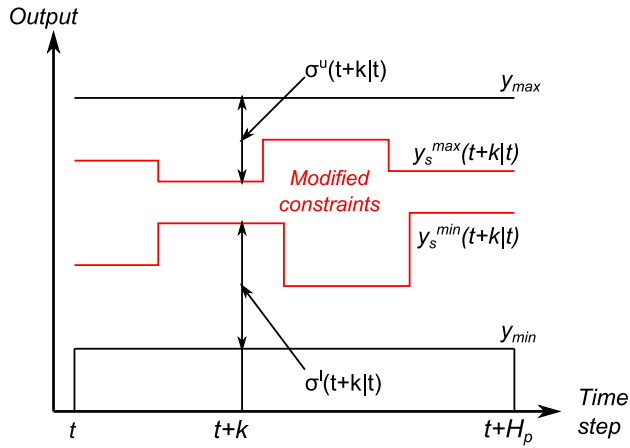

Fig. 4. Safety zones.

In (Brdys and Ulanicki, 1995), hard limits on tank capacities in a drinking water distribution system were additionally reduced to an MPC optimization task by introducing so-called safety zones. This is illustrated in Figure 4, where $\sigma^u$, $\sigma^l$ denote the upper and lower safety zones modifying the original upper and lower limits $y^{max}$ and $y^{min}$ constraining the output to produce the modified output constraints to be used in the model based optimization task of MPC. The safety zones were determined large enough in order to compensate uncertainty in the water demands so that the model based optimized control actions satisfied the original tanks constraints when applied to the physical water distribution system. Replacing in the MPC model based optimization task the original state/output constraints with a set of more stringent constraints which preserve feasibility for any scenario of uncertainty in the system model dynamics is a key idea of this constraint restriction approach. A disturbance invariant set was designed a priori in (Chisci et al., 2011) to produce suitable restrictions of the constraints for linear systems. The conservatism of methods based on the invariant sets and difficulties in calculating these sets for nonlinear system dynamics impose serious limitations on applicability of these methods.

In (Brdys et al., 2011) the safety zones were derived for nonlinear constrained dynamic networks to achieve not only robust but also recursive feasibility of the MPC. The numerical algorithm was proposed to calculate the safety zones explicitly off-line based on the uncertainty prediction error bounds and utilizing Lipschitz constants of the nonlinear network mappings. A generic approach to synthesise RFMPC that utilizes the safety zones, which are iteratively updated on-line based on the MPC information feedback was proposed in (Brdys and Chang, 2002a) and applied to the drinking water quality and hydraulics control in (Duzinkiewicz et al., 2005; Brdys and Chang, 2002; Tran and Brdys, 2013) and to integrated wastewater systems in (Brdys et al., 2008). The robustly feasible model predictive controller with iterative safety zones is practically applicable to nonlinear systems and the conservatism due to the uncertainty is much reduced as the safety zones are updated on-line utilizing the measurements from the real system over the prediction horizon. The recursive feasibility is guaranteed by selecting the prediction horizon long enough.

The controller structure is illustrated in Figure 5. The control inputs are produced by solving the model based optimization task, where the unknown disturbance inputs over the prediction horizon are represented by their updated predictions and other stationary uncertainty factors are replaced by their estimated values, for example by Chebyshev centers of the set membership estimates. Hence, the MPC optimization task is deterministic and therefore computationally less demanding.

Moreover, the original state/output constraints are modified by the safety zones provided by the Safety Zones Generator. The initial state in the output prediction model is taken directly from the plant measurements or it is estimated using these measurements. The control inputs are then checked for robust feasibility over the prediction horizon. First, a robust prediction of the corresponding plant output is produced in terms of two envelopes and bounding a region in the output space where the plant output trajectory would lie if the inputs were applied to the plant. The plant model with complete set bounded uncertainty description is utilized to perform the robust output prediction. The robust feasibility is now verified by comparing the envelopes robustly bounding the real (unknown) output against the original output constraints.

Determining the robustly feasible safety zones is done iteratively and typically, a simple relaxation algorithm is applied to achieve it. In order to achieve sustainable (recursive) feasibility of the RFMPC the safety zones are iteratively determined on-line over the whole prediction horizon and this is still computationally demanding. In a recent work (Brdys et al., 2011) the safety zones were applied to parsimoniously parameterize recursively feasible invariants sets in the state space and a computational algorithm was derived to calculate off-line the zones and the invariant sets. An operational computation burden of the resulting RFMPC was then significantly reduced at the cost of an increased conservatism of the control actions produced and consequently the increased suboptimality of MPC. Clearly, the safety zones and invariant sets are recalculated when a prior uncertainty bound changes. A rigorous mathematical analysis of the convergence of the iterative algorithms suitable to calculate the safety zones was performed. The problem was formulated as of finding a fixed point of a nonlinear mapping. A simple relaxation algorithm was derived as well compromising between the number of iterations requiring measurement feedback from the plant and the calculation complexity.

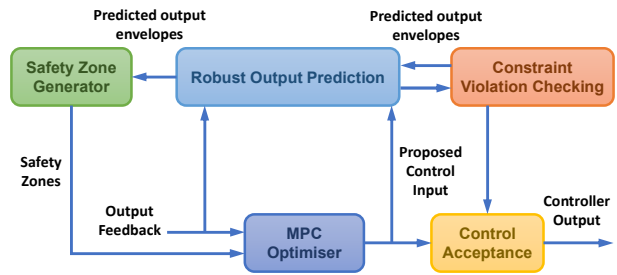

Fig. 5. Structure of RFMPC with iterative safety zones.

*3.2 Softly Switched Robustly Feasible Model Predictive Controller (SSRFMPC)*

The operational states (OS) were introduced in order to capture different operational conditions (Brdys et al., 2008). A current operational state (OS) of P&L is determined by the states of all the factors which influence the P&L ability to achieve the prescribed control objectives. These include: states of the P&L processes; states of the sensors, actuators and communication channels (e.g., faults), states of process anomalies, technical faults, current operating ranges of the processes, states of the disturbance inputs.

The typical operational states are: normal, disturbed and emergency. Not all control objectives can be satisfactorily achieved at a specific OS. This is identified by performing an adequate a prior analysis. Given the control objectives a control strategy capable of achieving these objectives is designed or chosen from the set of strategies designed a prior. In this way a mapping between the operational states and suitable control strategies to be applied at these OS can be produced. It should be pointed out that there can be more than one normal, disturbed and emergency operational states and they constitute the separated clusters in the OS space equipped with the links indicating transfer between the specific operational states. In a triple of ordered and linked of the normal, perturbed and emergency operational states, a deterioration of CIS operational conditions forces the P&L system CIS to move into the perturbed operational state. The control system is expected to adapt its current control strategy to the new operational state as otherwise the P&L CIS with not adequate control strategy in place can be further forced to move into the emergency operational state. Being safely in the perturbed operational state the agent senses and predicts changes in the current OS and if it moves back to the normal OS, for example, the intelligent agent starts adapting the control strategy back to the normal one.

Naturally, the control strategies are designed by applying the robustly feasible model predictive control technology.

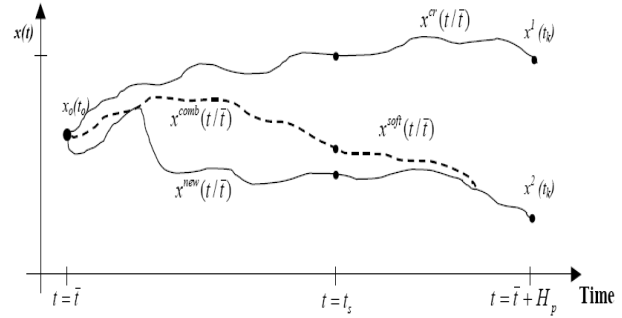

Fig. 6. Soft switching.

A hard switching from the current control strategy to the new one may not be possible due to at least two reasons. First, the immediate replacement in the control computer of the current performance and constraint functions by those defining the new control strategy may lead to the infeasible optimization task of the new strategy with the current initial state (Wang et al., 2005; Brdys and Wang, 2005). Secondly, very unfavorable transient processes may occur and last for certain time period as demonstrated in (Liberzon, 2003).

Alternatively, the switching can be distributed over time by gradually reducing the impact of the current (old) control strategy on the control inputs generated and strengthening the new control strategy impact (Fig. 6). The switching starting at $t = \bar{t}$ would complete at $t_s = \bar{t} + T_s$, where $T_s$ denotes duration time of the switching process. As opposed to the hard switching this is a soft switching. The soft switching was proposed and analyzed for linear constrained systems in (Wang et al., 2005; Brdys and Wang, 2005) and for the nonlinear systems in (Tran and Brdys, 2013). It was proposed to technically implement the soft switching by designing so called intermediate combined predictive control strategies in a form of a convex parameterization of the performance and constraint functions of the current (old) and desired (new) both strategies. Selecting on-line the parameters produces a sequence of the combined strategies and the new strategy is reached at the finite time $t_s$. A dedicated Supervisory Control Layer (SuCL) is introduced in order to identify on-line the OS's, initiate the switching process, manage its design and implementation

In (Wang and Brdys, 2006) an algorithm, which terminates the soft switching in a minimal time was proposed for linear constrained systems. The minimum switching time algorithm for nonlinear network systems was recently proposed in

(Tran and Brdys, 2013). The soft switching between hybrid RFMPC strategies was investigated for linear hybrid dynamics in (Wang and Brdys, 2006a) producing certain stability results. The soft switching was applied to optimizing control of integrated waste water treatment systems in (Brdys *et al.*, 2008) and to hydraulics control in drinking water distribution systems facing during their operation pipe bursts as well as sudden and lasting pressure increases, which would cause the pipe bursts if the normal operational strategies are maintained (Tran and Brdys, 2013). The RFMPC with not iterative safety zones (Brdys *et al.*, 2011) for generic nonlinear network systems was applied to design the control strategies for each of the OS. A recent research on truly Pareto multi-objective MMPC reported in (Kurek and Brdys, 2010) has produced results showing an enormous potential of the MMPC to develop new high dynamic performance soft switching mechanisms. There are still problems with performing on-line the computing needed to produce accurate enough representation of the Pareto front. Hybrid evolutionary solvers implemented on computer grids with embedded computational intelligence mechanisms that are designed based on fuzzy-neural networks with the internal states are investigated in order to derive more efficient solvers of the multi-objective model predictive controller optimization task.

*3.3 Cooperative Distributed SSRFMPC*

The softly switched robustly feasible model predictive control layer and the supervisory control layer are the functional layers in an overall multilayer structure of the reconfigurable autonomous agent capable of meeting the desired operational objectives under wide range of operational conditions. The complexity of the P&L may necessitate distribution of the operational tasks over a number of dedicated agents. Strong physical interactions exist during plant operation so that the agents must cooperate in order to successfully achieve the overall objectives. The desired multi-agent structure would be produced by suitable decomposition of the overall objectives to be followed by decomposition of the functional layers of a single global multilayer agent designed as above.

As the RFMPC is an optimization based technology then the well-known decomposition methods of the optimization problems can be applied to produce hierarchical structure of the RFMPC with the regional units and a coordinator integrating the regional actions. This would produce a hierarchical distributed multi-agent structure with minimized information exchange achieving an excellent operational performance due to the agent cooperation through the coordinator. This has not been done yet. The price coordination mechanism with feedback (Findeisen *et al.*, 1980; Brdys and Tatjewski, 2005) is very appealing. However, it needs to be further developed so that the robust feasibility of the actions generated by the distributed agents can be recursively guaranteed on-line for heavily state/output constrained systems, not only for the control input constrained system. Although the direct coordination mechanism does not suffer from this drawback its applicability is limited by the availability of efficient algorithms for solving difficult non- differentiable optimization tasks. However, an intensive research is in progress. Alternatively, developing not coordinated distributed RFMPC where the agent cooperation is non iterative and is performed by exchanging information about the most recent control/decision actions generated by the agents over the prediction horizon has attracted immense attention of the control community during the last decade (Chang *et al.*, 2003; Ding *et al.*, 2010; Dunbar, 2007; Dunbar and Murray, 2006; Zheng *et al.*, 2011, 2013; Venkat *et al.*, 2007; Zheng and Li, 2013). Excellent surveys can be found in (Scattolini, 2009; Rawlings and Mayne, 2009). The information exchanged is utilized by the agents to robustly predict the interaction inputs into the model-based optimization tasks of their RFMPC's.

Formulation of a distributed model predictive control architecture as a bargaining game problem allows each MPC subsystem to decide whether to cooperate or not depending on the benefits that the subsystem would gain from the cooperation (Valencia *et al.*, 2011). The resulting control system can be seen as an enhancement of the non-iterative distributed MPC based cooperative control. The required horizontal information exchange between the regional agents can be immense and certainly not acceptable by real life communication networks. The operational performance can be poor due to conservatism of the mechanisms of these distributed structures, which secure the feasibility. Finally, in order to achieve high operational performance in a cost effective manner under strong interactions the distributed agents must be coordinated.

Research on the hierarchical structuring the soft switching mechanism is in progress. The communication protocols implementing the information exchanges between the agents directly or through the coordinator require security features to be embedded in these protocols and beyond with a whole information system to be applicable. Although much work has been done for information systems the results are not directly applicable to the engineering systems, which require more control engineering system technologies rather than the computer science methods in place (Freggen *et al.*, 2005).

The decentralized follow-up control methods are applicable to structure the agent lowest layer for MAS purposes.

## 4. DISTRIBUTED AND COOPEATIVE APPROACHES IN PRODUCTION AND LOGISTICS

This section aims at overviewing existing paradigms and specialized cooperative technologies which are specially designed for the needs of production and logistics.

*4.1 Cooperative Engineering*

Cooperative Engineering is one of the great achievements of Enterprise Modelling. However, new factors, such as the fast evolution of information and communication technology (ICT) or the need to set up alliances among different types of enterprises, quickly, in order to benefit from market opportunities, are causing new types of problems, like interoperability, appeared in the Enterprise Modelling context. MES (Manufacturing Execution Systems) solutions provide real time information about what is happening in the shop floor, for managers (under a strategic approach) as well as for workers (under a purely operative approach). It is also an information bridge between Planning Systems used in Strate-gic Production Management (such as ERP – Enterprise Resource Planning) and Manufacturing Floor Control as SCADA (Supervisory Control and Data Acquisition). It links the Manufacturing Information System's layers (Strategic Planning and Direct Execution) through the adequate on-line managing and control of updated information related with the basic enterprise resources: people, inventory and equipment (Mejía, *et al.* 2007). The enormous importance acquired by MES resides, in a significant percentage, on its functionalities and their interaction with the compounding elements of the industrial plant environment. Core functions of MES include Planning System Interface, Data Collection, Exception Management, Work Orders, Work Stations, Inventory / Materials and Material Movement. MES supporting functions could include the following Genealogy, Maintenance, Time and Attendance, Statistical Process Control, Quality Assurance, Process Data and Documentation Management. However, there is an increasing need to provide support defining and implementing an interoperability relationship between these manufacturing software and business applications such as ERP systems (Panetto and Molina, 2008).

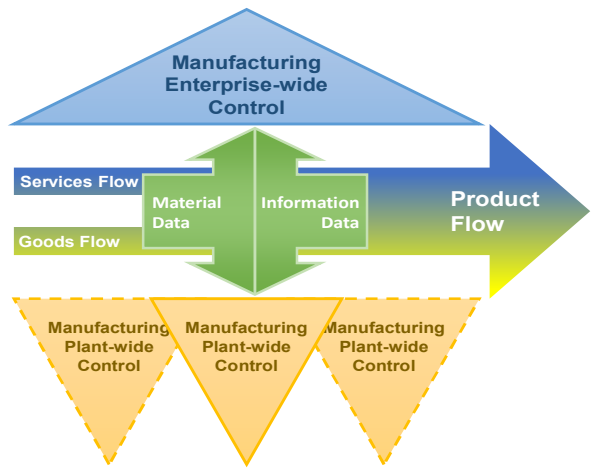

Fig. 7. Enterprise-wide control (Morel *et al.*, 2003).

In order to support the requested Business to Manufacturing (B2M) interoperation, the standard IEC 62264 (IEC, 2002) defines models and establishes terminology (semantics) for defining the interfaces between an enterprise's business systems and its manufacturing control systems. It describes the relevant functions in the enterprise and the control domain and the objects normally exchanged between these domains. It is becoming the accepted model for B2M integration and interoperability. In this context, the main modelling concept is to make the product interactive as the 'controller' of the manufacturing enterprise's resources for enabling 'on the fly' interoperability relationships between existing product-systems and ensuring coherence between the physical and information flows all through the product life-cycle (Figure 7) (Morel *et al.*, 2003; Panetto *et al.*, 2007).

The following research issues are considered challenging for the next years to come:

- Enterprise architecture needs addressing more on how to align of business strategy to technology for implementation, and not just focused on business or IT with separated research and development
- It is necessary to develop an Enterprise architecture language at a high level of abstraction for representing enterprise architectural structure, characteristics and properties at early stage of design.
- Existing architecture design principles and patterns were not developed to a satisfactory level to allow bringing significant improvement to enterprise architecting. More research is also needed in this area to promote the reuse of good practices and theories.
- The development of an ontology precisely defining concepts and properties of enterprise architecture domain is challenging. This ontology is needed to allow a clear understanding of the universe of discourse in this domain and avoid multiple and sometimes redundant developments of architectural proposals. Enterprise architecture ontology also contributes to semantic interoperability between different enterprise architecture proposals (Whitman and Panetto, 2006).

### 4.2 Holonic Production Control

The origins of holonic systems are insights in complex-adaptive systems theory (Waldrop, 1992) and bounded rationality (Simon, 1969). The concept of an *autocatalytic set* calls for maximizing the (critical) user mass. Bounded rationality calls for non-monolithic designs exhibiting time-varying loose hierarchical structures.

The PROSA architecture answers the latter by supporting time-varying aggregation of holons into larger holons. Critical user mass cannot be achieved by research prototypes (that requires actual industrial deployment) but their design delineates the maximum size of such user mass. PROSA divides the system into components in a manner that maximizes this potential for the user mass size.

To this end, PROSA cleanly separates the resources from the activities that use these resources to manufacture products. It also separates the managing of the logistic aspects (product routing, processing step assignment to resources) from the technological aspects (which sequences of processing steps are valid, which resources are capable of which operation). In combination with the support for aggregation, the ratio of user mass over the complexity of the holons is optimized.

PROSA has yet another property to guarantee it can handle challenges that present themselves: a structural reality-mirroring decomposition. A PROSA cooperative control system comprises a mirror image of the production system, tracking its changes and reconfigurations and connecting its components in manners that also reflect reality. This provides unlimited scale-ability (at least in principle). Indeed, it builds a mirror image of something that already scales up to the size of our universe. Note that any kind of functional or role-based decomposition is an inferior choice in this respect.

A much-ignored property of PROSA is the price it pays to achieve the above: *unfinished business*. PROSA leaves most of the design work, needed to develop a cooperative control system, to the implementers of an actual system. It only is a reference architecture not even a system architecture. Work-by-others has produced architectures that are partial instantiations of PROSA such as ADACOR (Leitao, 2006) or (Sallez, 2009; Zambrano, 2013; Pach, 2014).

As PROSA needs additional development to produce a usable control system, researchers have elaborated designs that include some concrete distributed and decentralized decision-making mechanisms; in contrast, PROSA keeps all options open by not providing or specifying anything. In particular, researchers have investigated market-based designs.

The main issue with market-based designs is myopia and/or the combinatorial explosion of its straightforward solutions (Zambrano *et al.*, 2013). As Parunak stated in a discussion during an AAMAS conference, a market mechanism forces the use of a utility function that reduces a multi-dimensional complex reality into a single scalar coupled to a fully-instantiated choice (service to be delivered at this price). That is an enormous loss of information and it is incapable of including a complex collection of conditional future commitments that will impact this utility. As a consequence, market based designs had successes where this myopia is not an issue; typically in systems that return to reference state after every action on which the market decides. But more complex situations require an effective look-ahead and the ability to make agreements/commitment regarding future actions and allocations.

The application of machine learning has a strong prospective in this matter. Stochastic, distributed resource allocation problems, with a special focus on production control, based on RL agents was analyzed by Csáji and Monostori (2008).

Alternative manners to address this myopia are also discussed below in sub-sections 4.3 and 4.8.

*4.3 Indirect Cooperation*

The above-discussed PROSA design supports an SSOT (single source of truth) design, but fails to preserve this until there is a working control system. It is like a map in navigation, useful but it still needs a navigator that can read this map and generate routing instructions. Note that SSOT is a highly desirable property in any system design. For instance, software and data maintenance only needs to look at the single affected element for every change in the corresponding reality (when a bridge is destroyed only the corresponding element of the map needs adjustment).

A shortcoming of the basic PROSA design is that only local information is available. There is no information related to facts that are remote in space or time. Indirect cooperation mechanisms are capable of delivering such information service without forfeiting SSOT (at the cost of a small time delay). Such a mechanism is the delegate multi-agent system or D-MAS (Holvoet 2010; Valckenaers 2005).

The information handling offers no guarantee that non-local information remains valid. A possible solution detected by an ant agent may be invalidated when another order holon reserves its slot on a resource (e.g., because it has priority). To cope with this issue, the design implements *forget and refresh* mechanisms. Information (reservations in an agenda) has a limited time span and needs to be refreshed regularly.

This way, the design copes with the dynamics in its world-of-interest and changes caused by decision-making subsystems.

Other indirect cooperation mechanisms exist. Some will reflect facts, physical or mental states, for instance indicating the presence of a batch at some point and place in the future that other orders may want to join. Another example is to indicate the predicted position and time of the system bottleneck based on a given criterion. Others will represent choices and, for instance, will attract or repel. The main difference between reality-reflecting (including intentions and commitments) and decision-making mechanisms is the compose-ability. The former has no issues whereas the latter requires the design to resolve conflicts regarding authority over actions and resource allocations.

*4.4 Dynamic Scheduling and Real-time Assignment*

Scheduling is the process of assigning tasks to a set of manufacturing or logistic resources with the objective to optimize a criterion taking into account task precedence constraints, limited resource capacities, task times and release dates for products.

Most of theoretical scheduling approaches largely ignored the dynamic character randomness of production and logistic systems (Weirs, 1997; Pinedo, 2002). Nevertheless, in manufacturing and logistic environments, unexpected events arise and so forces modifying the schedule (Stoop et al., 1996; Cowling et al., 2002; Viera et al., 2003). Unexpected events are, for example: machine breakdowns, tool failures, unavailability of tools or employees, shortage of raw material or components, defective or inadequate material or components, modifications of deadlines, order cancellations, late arrivals of orders and changes in manufacturing processes, etc. Thus, a schedule often becomes outdated before the moment it is finished.

Some authors discussed the gap between scheduling theory and the needs of manufacturing systems and logistics (MacCarthy et al., 1993; Cowling et al., 2002). Taking into account this situation, in the current research a large place is devoted to dynamic scheduling and real-time assignment techniques (Dolgui and Proth, 2010).

The competitive market encouraged by powerful data processing, communication systems and international trade agreements, has affected the structure of production and logistic systems, necessitating integration of all the activities as well as requiring flexibility with regard to market changes. Thus, in nowadays production systems the objective is to schedule and reschedule tasks online. Therefore, the most important perspective is in developing methods for real-time assignment of tasks to resources being able to reschedule "online" the whole supply chain in case of unexpected events and to react immediately to customers' demand (Chauvet et al., 2000; Dolgui and Proth, 2010).

*4.5 Cooperative Scheduling*

The above-discussed holonic production control leaves the exact nature of the decision-making open to the developers of an actual system. Among the possibilities, there is the option

to cooperate with a scheduler (Verstraete, 2008; Novas, 2012; Van Belle, 2013). This involves that:

- The first D-MAS, exploring for solutions, dedicates a significant percentage of its efforts (of its ants) to virtually executing routings that comply with the externally provided schedule. Note that, where needed, this virtual execution must handle actions that are not covered by the scheduler (e.g., transport by an AGV).
- The selection criterion, used by the order holon, for the preferred solution must favor solutions that follow the external schedule, provided their performance is in line with the schedule.
- The local agenda-managing policies of the resource holons give priority to visits in compliance with the external schedule.

Obviously, there remain many aspects to be investigated when implementing such scheme. Noteworthy is that the short-term prediction capability of the holonic control allows to employ schedulers that require longer computation times when they are initialized with the predicted state for the time when their results will be available.

*4.6 Bucket Brigades*

An example of self-organizing production systems is bucket brigades (Bartholdi and Eisenstein, 1996; Dolgui and Proth, 2010). For such an assembly line, each worker moves with the product while working. As soon as the last worker completes the product, he/she walks back upstream to take over the work of the predecessor, who then goes upstream to free up the first worker, who then moves to the beginning of the assembly line and starts work on a new product.

The most important advantages of bucket brigades are:

- It naturally redistributes the workload among workers depending on their efficiency.
- The flow of products is self-organizing, there is no centralized management.
- The obtained assembly line is agile and flexible, it adapt quickly to unexpected events.
- Work in progress is minimal, quality is improved.

A survey on bucket brigades and their industrial applications is given in (Bratcu and Dolgui, 2005), a simulation study is presented in (Bratcu and Dolgui, 2009).

*4.7 Production Networks and System Integration*

Systems Integration is generally considered to go beyond mere interoperability to involve some degree of functional dependence. While interoperable systems can function independently, an integrated system loses significant functionality if the flow of services is interrupted. An integrated family of systems must, of necessity, be interoperable, but interoperable systems need not be integrated. Integration also deals with organizational issues, in possibly a less formalized manner due to dealing with people, but integration is much more difficult to solve, while interoperability is more of a technical issue. Compatibility is something less than interoperability. It means that the systems/units do not interfere with each other's functioning. But it does not imply the ability to exchange services. Interoperable systems are by necessity compatible, but the converse is not necessarily true. To realize the power of networking through robust information exchange, one must go beyond compatibility. In sum, interoperability lies in the middle of an "Integration Continuum" between compatibility and full integration. It is important to distinguish between these fundamentally different concepts of compatibility, interoperability, and integration, since failure to do so, sometimes confuses the debate over how to achieve them. While compatibility is clearly a minimum requirement, the degree of interoperability/integration desired in a joint family of systems or units is driven by the underlying operational level of those systems (Panetto, 2007).

*4.8 Autonomous Logistic Processes*

The design of the holonic production control system has been translated to logistic execution systems (Van Belle, 2013). The overall design could be used without modification. The need to cooperate with a scheduler, or other mechanisms to guide the search for good solutions, is higher because the search space is huge and comprises lots of very poor solutions. The need to handle multi-resource allocation is also more prominently present. However, this does not affect the basic design while the improvements and enhancements are relevant for production control (cross-fertilization).

The advantages of a holonic Logistics Execution System (LES) comprise the ability to use simpler schedulers (in the software development and in the computational complexity sense). More importantly, the presence of order holons (mirroring real-world activities) connecting the resource holons represent a major opportunity for system integration, networked production and multi-hop logistics. A major pitfall when attempting to integrate systems into larger systems by integrating the resources while capturing activities in data formats is that these format standards and specs are:

- Either too simplistic and unable to cope with the complexity of the world-of-interest
- Or too expressive (i.e. tend to become a full-fledged scripting and programming language) for the user mass and economic support that they may gather.

Integration will fail or result in poor performance; there is interoperability but the common denominator, which is the upper bound of what interoperability may achieve, is unsatisfactory.

*4.9 Collaboration in Supply Chains*

Collaboration issues across the supply chain were stressed in (Chung and Leung, 2005). Other researchers, for example (Barbarosoglu, 2000; Zimmer, 2002), considered the two-echelon models of buyer–vendor systems with the idea of joint optimization for supplier and buyer. A three-echelon model that includes the manufacturer, distribution center and retailer was suggested in (Kreng and Chen, 2007).

Indeed, as mentioned in a large number of publications, for example (Sterman, 1989; Blanchard, 1983), there is a distortion of demand (bullwhip effect) when moving upstream in a supply chain. A possible remedy deals with close collaboration of the manufacturer with the retailer. In (McCullen and Towill, 2001), the authors suggest linking factory plans to real-time customer demand. These approaches are known as methods based on information transparency or supply chain visibility.

The advantage of sharing information among the different nodes of the supply chain and generalize the concept of collaboration between the nodes of a supply chain were emphasized in (Dolgui and Proth, 2010). Some models and simple strategies illustrated with simulation were presented, especially to show the benefits of collaboration and information sharing. These studies demonstrated that the bullwhip effect can be reduced drastically in the case of collaboration and information sharing.

## 5. CASE STUDIES

Here two case studies are presented. The first one is about generating short-term forecasts by means of D-MAS, while the second one deals with the application of distributed RFMPC to a drinking water distribution system.

*5.1 Short-term Forecasts by D-MAS*

The knowhow concerning holonic manufacturing execution systems, which is PROSA-based and generates short-term forecasts by means of D-MAS, has been transferred to industry (Valckenaers and Van Brussel, 2005; Holvoet et al., 2010). This transfer occurred through the development of a prototype implementation for additive manufacturing.

The industrial partner in the additive manufacturing domain employs an in-house custom MES because commercially available solutions, benefiting from a sound user community, lack the proper functionality. In particular, the three-dimensional nesting, which requires domain-specific matching/grouping and ungrouping, and process variability could not be handled by a COTS solution.

This in-house MES did not support the short-term self-organizing prediction functionality of holonic manufacturing execution systems (Valckenaers et al., 2011). The development implemented this forecasting capability as an add-on to the existing in-house MES. This mainly consisted of developing the required executable models mirroring the world-of-interest (i.e. the additive manufacturing processes).

The holonic MES generates short-term forecasts by virtually and repeatedly executing the envisaged of product routings and processing steps using the above-mentioned models (Valckenaers and Van Brussel, 2005). In additive manufacturing, a high-powered laser scans a material surface to build – layer by layer – a product that is entirely defined by the data driving the laser scans. The material typically is a liquid polymer that gets solidified when the laser beam passes over it, or it is a metallic powder whose grains are melted together by the heating from the laser beam. Originally, this technology was used for rapid prototyping but increasingly finished products are made through additive manufacturing. An important market is the medical world where implants (e.g., in titanium) or surgical fixtures (e.g., that will guide instruments during brain surgery) are welcomed.

To ensure productivity, a machine tool will not build a single product – layer by layer – but software will be used to combine multiple products within the work space of the machine. This is called nesting. More precisely, it is a 3D nesting problem where, e.g., sheet metal laser cutting corresponds to a 2D nesting problem. This is vital for the manufacturing organization as production times depend foremost of the number of layers and somewhat less on the particular laser scanning pattern.

The generation of forecasts through virtual execution therefore involves solving this nesting problem, which includes the selection of products (shapes) that will be produced together and the position of the selected products within a machine's workspace. In practice, this nesting optimization involves time-consuming computations and, often, human intervention. As a consequence, a specific challenge during the development of the required executable models was the modeling of these three-dimensional nesting mechanism.

The base design of the holonic execution system had to be enhanced by supporting models that approximate these nesting procedures without requiring those time-consuming calculations or human intervention when refreshing (recall that a D-MAS employs a forget and refresh mechanism) or in case of minor changes in the (predicted) situation. If these approximations produce inaccurate data, the holonic execution system will handle it as a disturbance, which is anyhow a core functionality of this holonic system.

The technology transfer project successfully developed a prototype, connected to the in-house MES that generated these short-term forecasts. Through its design, this combination of two cooperating systems is capable of sharing and propagating these forecasts along the supply lines, thus enabling a proactive coordination with the customers. For instance, surgeons that need custom fixtures to perform an operation requiring accurate positioning may plan and organize their work with less slack time.

From a practical implementation perspective, the academic prototype software had been developed in Java whereas the in-house MES used C# and .NET technology. After some initial discussions, the project decided to keep both technologies and establish a communication link to achieve the required cooperation. This was the situation in the early phase of the technology transfer project.

At a later point in time, when work on this link was about to start, the holonic execution systems technology had been implemented in Erlang/OTP within another project (EU project MODUM), where this implementation incorporated the latest developments, was significantly better-performing and more stable (Erlang was designed to develop scalable, distributed and very robust systems). The team decided to check whether was possible, with very little effort, to switch to this Erlang version.

Within one day, the team established a communication link between the in-house MES and the Erlang version of the holonic systems software. This triggered the decision to switch to the Erlang version, which required a couple of weeks. This Erlang version successfully demonstrated its capability to generate short-term forecasts in cooperation with the in-house MES.

*5.2 Distributed Robustly Feasible Model Predictive Control in Drinking Water Distribution Systems*

*5.2.1 Introduction and Problem Statement*

Drinking water distribution system (DWDS) delivers water to domestic users. Hence, the main operational objective is to meet for every consumer the water demand of required quality (Brdys and Ulanicki, 1995). For safe and efficient process operation, monitoring and control systems are needed. In this section the monitoring system is assumed in place and the control system for DWDS is pursued. There are two major aspects in control of drinking water distribution systems (DWDS): quantity and quality. The quantity control deals with the pipe flows and pressures at the water network nodes producing optimized pump and valve control schedules so that water demand at the consumption nodes is met and the associated electrical energy cost due to the pumping is minimized (Brdys and Ulanicki, 1995; Boulos, *et al.,* 2004).

Maintaining concentrations of water quality parameters within prescribed limits throughout the network is the main objective of the quality control system. In the section, only one quality parameter is considered: the free chlorine concentration. Chlorine is the most common disinfectant used in DWDSs worldwide. It is not expensive and effectively controls a number of disease-causing organisms. As the chlorine reactions with certain organic compounds produce disinfectant by-products (DBP) THM compounds that are health dangerous (Boccelli, et. al 2003) the allowed chlorine residuals over the DWDS are bounded above. Hence, the operational objective of maintaining desired water quality is expressed by certain lower and upper limits on the chlorine residuals at the consumption nodes. The available water quality sensor measurements over DWDS are very limited so that the quality state must be estimated for monitoring and control purposes (Langowski and Brdys, 2007). Recently, a comprehensive mathematical model of water quality was developed (Arminski *et al.*, 2013) and applied to derive a chlorine and DBP dynamics model suitable for the robust estimator design utilizing a cooperative property of the model dynamics (Arminski and Brdys, 2013). The chlorine residuals are directly controlled within the treatment plants so that the water entering the DWDS has the required prescribed residual values. However, when travelling throughout the network the disinfectant reacts and consequently its major decay may occur, so that a bacteriological safety of water may not be guaranteed particularly at remote consumption nodes. Therefore, post chlorination by means of using booster stations located at certain intermediate nodes is needed. A problem of placement of the booster stations over a DWDS was investigated in (Prasad *et al.*, 2004; Ewald *et al.*, 2008) and some solution methods based on multi-objective optimization were provided. The chlorine residuals at the nodes representing outputs from the treatment plant and at the booster station nodes are the direct control variables for the quality control. Electricity charges due to pumping constitute the main component of the operational cost to be minimized. As there is an interaction between the quality and quantity control problems due to the transportation delays when transferring the chlorine throughout the network, a proposal to integrate these two control issues into one integrated optimization (control) problem was presented in (Brdys *et. al.*, 1995a) and a receding horizon model predictive control technique was applied to the integrated quantity and quality in DWDSs. Several solutions to the MPC optimization task were proposed applying the genetic search (GE) (Ostfeld *et al.*, 2002), mixed integer linear (MIL) algorithm (Brdys *et al.*, 1995a), sequential hybrid GE-MIL approach (Trawicki *et al.*, 2003) and nonlinear mathematical programming approach (Sakarya and Mays, 2000).

*5.2.2 A Single Agent - Centralized Two Time Scale Hierarchical Controller*

Due to different time scales in the hydraulic variations (slow) and internal chlorine decay dynamics (fast) the integrated optimization task complexity did not allow applying the integrated control to many realistic size DWDSs. While the hydraulic time step is typically one hour, the quality time step is for example five minutes and the prediction horizon due to tank capacities is typically 24-hour, the dimension of the optimization problem largely increases even for small-scale systems (Brdys *et al.*, 2000; 2013).

The *optimizing controller* at the *upper control level* (UCL) operates at the hydraulic slow time scale according to a receding horizon strategy. At the beginning of a 24-hour time period the DWDS quantity and quality states are measured or estimated and sent to the integrated quantity and quality optimizer. The consumer demand prediction is also sent to the optimizer. The simplified quality model assumes the same time step as the quantity dynamics model. Hence, the problem dimension is vastly reduced but the quality modelling error is significantly increased. Hence, solving the integrated quantity-quality optimization problem produces the optimized chlorine injection schedules at the booster and the treatment plant output nodes having poor quality. As the quality outputs do not influence the hydraulic variables (the interaction between quality and quantity is only one way from the quantity to the quality) the achieved optimized pump and valve schedules are truly optimal. Hence, the pump and valve schedules are applied to the DWDS and maintained during so called control time horizon, e.g., 2 hours. The quality controls need to be improved and this is performed at the *lower correction level* (LCL) by the fast *feedback quality controller* operating at the quality fast time scale. It samples the chlorine residual concentrations as it is required by its decay dynamics, e.g., with one minute sampling interval. In order to take advantage of the allowed quality bounds the centralized RFMPC with output constraints and the iterative safety zones was applied by (Brdys and Chang, 2001). A suboptimal approach is to specify a reference trajectory lying within the prescribed quality bounds and apply an adaptive indirect model reference controller to track this reference trajectory (Polycarpou *et al.*, 2001). The distributed RFMPC was applied at LCL for the first time in (Chang *et al.*, 2003). The single agent with centralized MPC with full hydraulics and quality information feedback achieving robust constraint satisfaction by fixed safety zones was applied for the first time to the integrated quantity and quality control problem in (Drewa *et al.*, 2007) and it is presented in section 5.2.3. The multiagent structure and algorithms for the two time scale hierarchical controller with RFMPC at both control layers are under development.

*5.2.3 Application to Gdynia DWDS Case-study*

A skeleton of the DWDS at Gdynia is illustrated in Figure 8 and its data are as follows: 3 underground water sources, 4 tanks and 3 reservoirs, 10 variable speed pumps, 4 control valves, 148 pipes, 134 pipe junction nodes, 87 demand nodes, 5 booster stations allocated at the quality control nodes, 129 quality monitoring nodes.

The accuracy of provided on-line demand prediction over 24-hour period was 5% for the first 10 hours and 10% for the remaining time slot of the 24-hour prediction horizon. The electricity tariff during 6 am-12a m and 3 pm-9 pm was $\eta = 0.12$ [$/kWh] and $\eta = [0.06$ $/kWh] during 10 pm-5 pm.

The DWDS skeleton is a simplified structure of the real one composed of such aggregated representations of the real system components that such system structure approximation remains viable for control purposes.

The centralized MPC controller was applied with the 2 hours hydraulic time step and 9 minutes quality time step. The results are illustrated in Figure 9 (resulting quality) and Figure 10 (resulting quantity). Comparison in the Figure 10 of the trajectory of a selected tank in Witomino, which is currently achieved at the site with the trajectory forced by the MPC actions, shows a very conservative operation of the current system. Such operation leads to high operational cost due to the electricity charges. It is implied by unavoidable difficulties in meeting the inequality constraints in this strongly interconnected system. The MPC utilizes the available tank capacity much better than the current operational strategy. An excellent quality control result is illustrated in Figure 9. The chlorine concentration in a junction node lies within the prescribed limits and it gets close to the lower limit, hence assuring limited production of harmful components due to the reactions of the organic matter with free chlorine.

The distributed RFMPC with cooperative agents will be applied in this section to derive the lower level controller with fast feedback from the quality measurements for the control architecture presented in Figure 12. The benchmark structure is illustrated in Figure 11. There are 16 network nodes, 27 pipes and 3 storage tanks in the system. All tanks are the switching tanks (pressure driven) and they can only be operated in a repeated sequential filing and draining cycles. The water is pumped from the sources (node 100 and node 200) by two pumps (pump 201 and pump 101) and is also supplied by the pressure driven tanks (node 17, 18, 19).

Nodes 16 and 8 are selected as monitored nodes as they are the most remote nodes from the sources. Hence, if the chlorine concentrations at these two nodes meet the quality requirements, then these requirements are also met at all other nodes over the DWDS. The chlorine concentrations at these nodes are the two plant-controlled outputs $y_1(t)$ and $y_2(t)$, respectively. There are two *quality control nodes*, where the chlorine is injected, in order to control the chlorine concentrations at the monitoring nodes: nodes 5 and 10. The booster stations are installed at these nodes as the actuators to produce the required chlorine concentrations $u_1(t)$ at the node 5 and $u_2(t)$ at the node 10. These are the quality control inputs and the controlled output in this DWDS benchmark. The fast feedback quality controller operates under the node 5 and $u_2(t)$ at the node 10. These are the quality control inputs and the controlled output in this DWDS benchmark. The fast feedback quality controller operates under the pump control inputs determined by the upper level controller as it is shown in Figure 13. Hence, the flows are determined. The RFMPC output prediction and control horizons are 24-hour while the quality control step is 5 minutes. Thus, the 24-hour control horizon is converted into 288 discrete time steps.

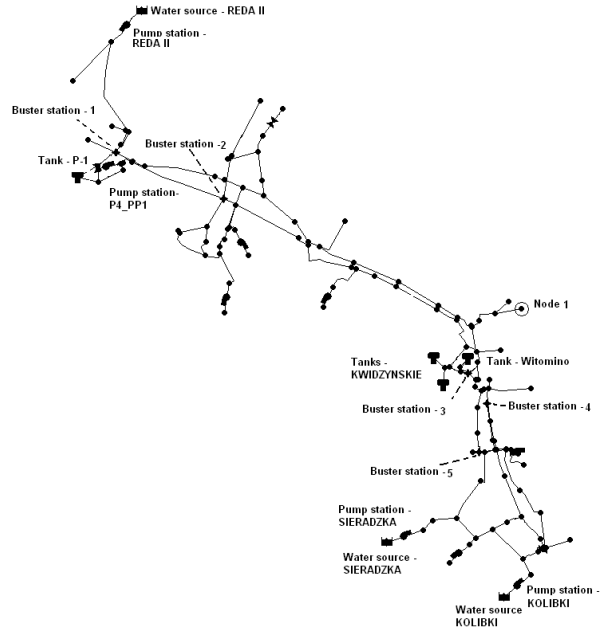

Fig. 8. A skeleton of the DWDS at Gdynia.

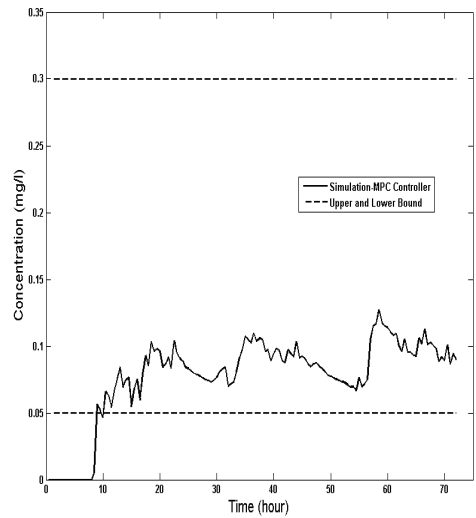

Fig. 9. Chlorine concentration in the quality monitoring.

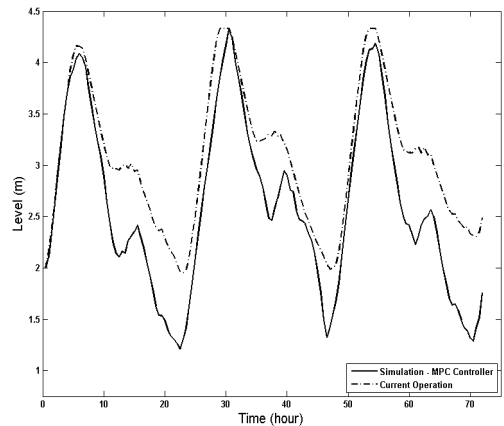

Fig. 10. Witomino – Tank level.

*5.2.4 Application of Distributed RFMPC with Cooperative Agents to Quality Control in DWDS*

The network is divided into two interacting zones. Each zone is controlled by the associated RFMPC agent. The agents cooperate by exchanging information about the most recent control/decision actions generated by them over the prediction horizon. This information is used to predict their interaction inputs in the model based optimization tasks. For the comparison purposes the performance of the centralized RFMPC is illustrated in Figure 14.

The control operational objectives are: to maintain the prescribed chlorine concentrations at the monitored nodes under the constraints on their instantaneous values with prescribed values at the end of the prediction horizon and meeting the actuator constraints due to the limits on the instantaneous values of the chlorine injections and their rate of change, which are prescribed in terms of bounds. The distributed RFMPC (DRFMPC) controller performance is illustrated in Figures 13 showing that the objectives are successfully achieved.

Comparing the results illustrated in Figure 13 with the results shown in Figure 14, especially during the time period from step 200 to time step 288, it can be seen that the control inputs are quite different. The injection at node 10 of the DRFMPC controller is more intensive than that of the centralized RFMPC. In the latter case, the control loop of the node 10-8 pair receives more chlorine contribution from the loop of the node 5-16 pair. In spite of the cooperation between the RFMPC agents of the DRFMPC the loop coordination is weaker. Hence, a compensation of the 'missing injection' is needed in order to achieve a comparable performance. This can only be done by the second control agent. In this DWDS case study, such ability to compensate a weakening of the coordination between local controllers is still within the capacity of this agent. Hence, the output constraints are still kept within prescribed limits.

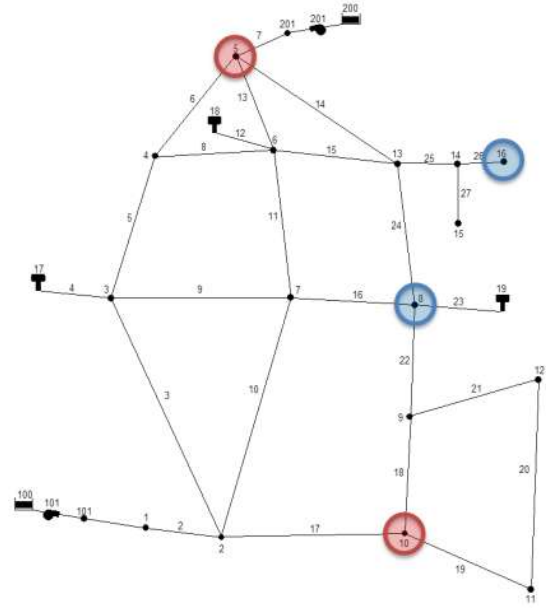

Fig. 11. Structure of DWDS benchmark.

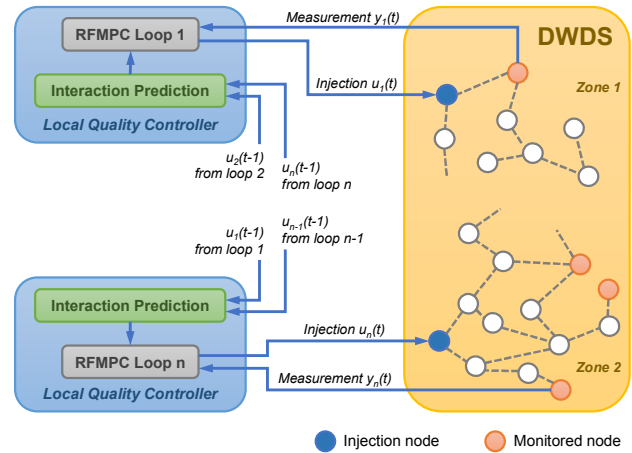

Fig. 12. Structure of distributed RFMPC.

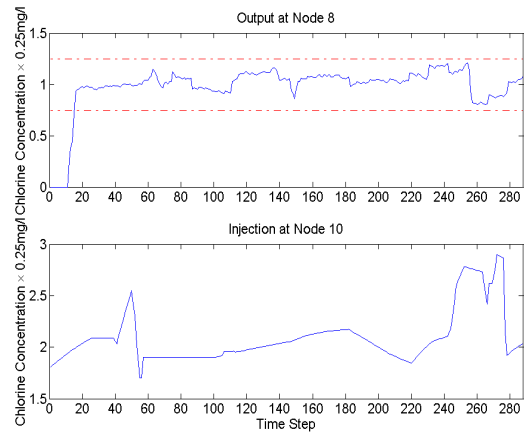

Fig. 13. DRFMPC: $y_1$ and $u_1$.

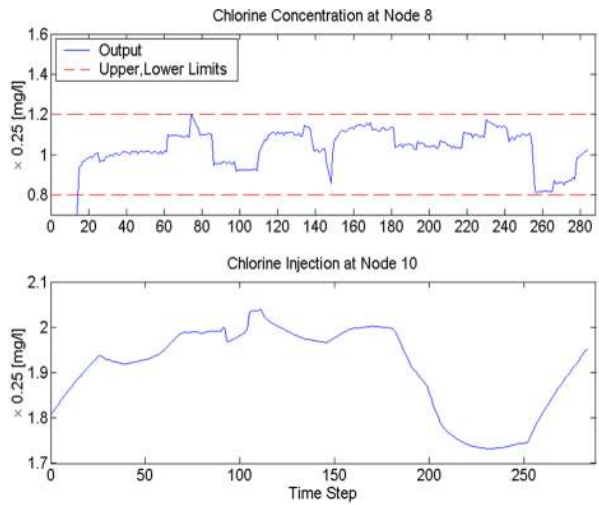

Fig. 14. RFMPC: $y_1$ and $u_1$.

## 6. CONCLUSIONS

Varying market demands, increasing volatility, abrupt changes, internal and external disturbances, as well as large number and variety of interconnected, interdependent entities call for a new control paradigm in production and logistics which can face these challenges and replace the traditional inflexible, pre-programmed, hierarchical control structures.

In the past several authors have argued that the future of manufacturing and logistics lies in network-like, dynamic, open and reconfigurable systems of cooperative entities.

The paper overviewed the advantages and disadvantages of such cooperative control approaches to production and logistics systems and surveyed results from information and communication technology (ICT) and control theory which can support developing such networks of cooperative entities.

Though there were considerable theoretical developments in related fields, such as control theory and ICT, and there are already some promising industrial applications of cooperative control, there are still many challenges to be faced when aiming for full-fledged cooperative production and logistics systems. These challenges include (a) decentralized, local information, and (b) limited processing capacities, which may result in (c) decision myopia; such cooperative system will need efficient (d) communication protocols and consensus mechanisms, which can also help (e) achieving high-level cooperation plans; the (f) security / confidentiality issues should also be taken into account as well as the potential of (g) chaotic, unstable behavior even if all the cooperating systems were stable. Addressing these may require developing new enterprise design principles, new architecture languages, ontologies, and applications of state-of-the-art results from several fields, such as control theory, ITC, cooperative game theory and distributed machine learning. On the other hand, these fields can also benefit from production and logistics, as they can provide many real-world problems with complex challenges to be solved.

ACKNOWLEDGMENTS

The authors from Hungary express their thanks to the Hungarian Scientific Research Fund (OTKA) for its support (Project No.: 113038).